\documentclass[prd,amsmath,nobibnotes,nofootinbib,superscriptaddress,preprintnumbers,twocolumn]{revtex4}

\usepackage{natbib}
\usepackage{amsmath,amsfonts,amssymb}
\usepackage{graphicx}
\usepackage{epstopdf}
\usepackage{type1cm}
\usepackage{eso-pic}
\usepackage{color}

\makeatletter
\AddToShipoutPicture{%
            \setlength{\@tempdimb}{.5\paperwidth}%
            \setlength{\@tempdimc}{.5\paperheight}%
            \setlength{\unitlength}{1pt}%
            \put(\strip@pt\@tempdimb,\strip@pt\@tempdimc){%
        \makebox(600,-800){\rotatebox{45}{\textcolor[gray]{0.90}%
        {\fontsize{4cm}{4cm}\selectfont{}}}}%
            }%
}
\AddToShipoutPicture{%
            \setlength{\@tempdimb}{.5\paperwidth}%
            \setlength{\@tempdimc}{.5\paperheight}%
            \setlength{\unitlength}{1pt}%
            \put(\strip@pt\@tempdimb,\strip@pt\@tempdimc){%
        \makebox(675,-950){\rotatebox{45}{\textcolor[gray]{0.85}%
        {\fontsize{1.cm}{1.cm}\selectfont{}}}}%
            }%
}

\begin{document}

%\shorttitle{}
%\shortauthors{Laughlin et al.} 

\title{Information Transmission
Between Financial Markets in Chicago and New York}

\date{\today}

\newcommand{\units}{\, \mathrm}

\author{Gregory Laughlin}
\affiliation{UCO/Lick Observatory, University of California, Santa Cruz, Santa Cruz, CA 95064, USA}

\author{Anthony Aguirre}
\affiliation{Department of Physics, University of California, Santa Cruz, Santa Cruz, CA 95064, USA}

\author{Joseph Grundfest}
\affiliation{Stanford Law School, 559 Nathan Abbott Way, Stanford, CA 94035}

\begin{abstract}

High frequency trading has led to widespread efforts to reduce information propagation delays between physically distant exchanges. Using relativistically correct millisecond-resolution tick data, we document a 3-millisecond decrease in one-way communication time between the Chicago and New York areas that has occurred from April 27th, 2010 to August 17th, 2012. We attribute the first segment of this decline to the introduction of a latency-optimized fiber optic connection in late 2010. A second phase of latency decrease can be attributed to line-of-sight microwave networks, operating primarily in the 6-11 GHz region of the spectrum, licensed during 2011 and 2012. Using publicly available information, we estimate these networks' latencies and bandwidths. We estimate the total infrastructure and 5-year operations costs associated with these latency improvements to exceed \$500 million.

\end{abstract}

\maketitle

\section{Introduction}

On September 1, 1949, AT\&T augmented its ``Long Lines" telephone service between New York City and Chicago with a 34-hop line-of-sight  4 GHz microwave radio relay. The 838-mile route traversed six states, and carried voice and television traffic between endpoint terminals located on AT\&T's New York City central office building, and the Congress Street Telephone Building in Chicago. The New York to Chicago segment was part of a larger 106-hop transcontinental microwave relay that linked Chicago to Denver and San Francisco, and which cost \$40 million to construct~\cite{LaFrance}. 

Historically, new technologies, such as the principles of radio communication underlying AT\&T's network, have been rapidly adopted for the advantage of participants in financial markets. From 1847 through 1851, Paul Reuter employed carrier pigeons between Brussels and Aachen to bridge a gap in telegraph stations on the route connecting Berlin to Paris, thereby providing a low-latency data feed for market-moving events~\cite{Reuter}. In 1865, the financier James Fisk completed a purpose-built telegraph line to Halifax in order to signal a fast steamship to cross  the Atlantic with instructions to short Confederate bonds at the close of the U.S. Civil War~\cite{Swanberg}.  One of Thomas A. Edison's best known inventions was the stock ticker, which transmitted market pricing information using the then newly practical transmission of information by telegraph.

In recent years, Moore's law has driven computerization of the financial exchanges, with an attendant decrease in trading latency, and a strong growth in automated, algorithmic trading. The speed of trading has  entered the special relativistic regime, in which speed-of-light related delays are a significant factor. Algorithmic strategies are now routinely controlled by computer servers that are physically co-located at the exchanges. Substantial effort and expense are invested in speeding up the trading pipeline; end-to-end latencies between order generation from strategy servers to order fulfillment and confirmation from the exchange matching engines are now measured in microseconds or less. Exchange data (such as the NASDAQ Itch 4.1 tick data analyzed here) are time-stamped to the nanosecond. Indeed, if latencies continue to decrease at the current exponential rate, it is a matter of a mere one to two decades before novel general relativistic effects associated with time dilation in the gravitational potential will begin to assume importance. For a further discussion of asset pricing in relativistic contexts, see~\cite{Haug}.

The relatively new practice of high frequency trading (ÒHFTÓ) is controversial and complex~\cite{sec10a}.  It is also pervasive, profitable, and secretive. HFT -- loosely defined as computerized trading of financial instruments, generally with the intention of holding instruments for short time periods, and subject to conditions computed by an algorithmic driver -- has been estimated to account for between fifty and eighty percent of executed volume on United States equity exchanges, even though it is practiced by fewer than two percent of all registered trading firms~\cite{lati,Poirier, McGowan}. Historically, high frequency traders have captured significant risk-adjusted returns~\cite{Brogaard}, but  industry-wide estimates of HFT revenues are highly variable and have been trending downward.

For 2009, considered a peak year for industry profitability, estimates ranged from \$7.2 billion to \$25 billion~\cite{McGowan,Tabb,Saraiya}.
A leading market analyst suggests that increased competition and lower trading volume have caused a significant decline in HFT industry revenues from \$7.2 billion in 2009 to \$1.8 billion in 2012~\cite{Tabb}. Other estimates suggest 2012 profits of no more than \$1.25 billion for the industry as a whole~\cite{Popper}. If we assume that HF traders  in aggregate capture a non-negligible fraction, $f_{c}$, of the minimum \$0.01 bid-offer spread on all US equity trades, then the yearly equity-trading profits, ${\cal P}$, for the  HFT industry, given an average traded daily volume, $V$, are $${\cal P}=f_c\left(\frac{V}{5\times10^{9}\, {\rm shares\, day^{-1}}}\right)\times\$12.5 \,{\rm billion}\,.$$ Typical daily volumes were $V={1\times10^{10}\, {\rm shares\,day^{-1}}}$ in 2009~\cite{Popper}, but have declined to $V \lesssim{5\times10^{9}\, {\rm shares\, day^{-1}}}$ by the second half of 2012\footnote{2012 volumes: http://www.batstrading.com/market\_summary/}. In addition, as estimated via trading volumes and minimum bid-offer spreads, equity options and equity futures trading supply additional revenues that sum to $\sim 1/3$ of those generated by stock transactions.
  
While academic research analyzing HFT has expanded substantially in recent years (see, e.g. bibliographies and literature reviews in~\cite{Smith,Biais}) details regarding the precise algorithms, trading strategies, and technologies deployed by high frequency traders, as well as the costs of implementing these strategies, remain scarce. 
This article draws on the quantitative analysis of publicly available data to strip away a portion of the secrecy that relates to two very specific forms of high frequency trading: equity market-making and liquidity-taking algorithms that rely on access to high speed information regarding futures market equity index price formation at a geographically distant exchange. The literature has long established that equity futures prices formed in Chicago lead cash prices formed in the US equity markets, which now, as a practical matter, reside in the exchange's matching engines housed at various data centers in suburban New Jersey. Many algorithmic traders with computers co-located at equity market trading facilities in New Jersey thus value low-latency access to price information emanating from Chicago, and in recent years, latencies dictated by the speed of light, both within and between exchanges, have become important. As detailed below, our analysis indicates that for the past two years, 20\% of all trading in the highly liquid SPY ETF can be specifically identified to occur in direct response to changes in the traded price of the near-month E-mini S\&P 500 futures contract.\footnote{It is interesting to note that of order $\sim10^4$ such trades occur daily; this is just $\sim2\times 10^4$ bits of information in the 6.5 hours  of the trading day, giving a data rate, $\langle  r \rangle \sim 1\,{\rm bit\, s}^{-1}$, not dissimilar to that of Edison's stock ticker.}

Our approach is twofold. First, we use the latency structure of inter-exchange communication -- as expressed in historical tick data -- as a quantitative tool to monitor the operation of a specific class of HFT algorithms. For 558 trading days, we correlate movements in the near-month E-Mini S\&P 500 futures contract, traded on the CME in the Chicago area, with response by the SPY and other equity order books at the NASDAQ and other New Jersey-based exchanges. As the CME orders are time-stamped to millisecond (ms) precision using the globally synchronized GPS-established time frame, we can measure end-to-end latencies and document high frequency cash equity market reactions to small exogenous futures market price shocks. 

Second, we document both the technology and the recent development of an ultra-low latency communication infrastructure that is used to transmit information between the market centers in the metropolitan Chicago and New Jersey/New York areas. 

We find that the cutting-edge method used to communicate financial information from Chicago to New Jersey has recently progressed through three distinct stages. Our analysis of the market data confirms that as of April, 2010, the fastest communication route connecting the Chicago futures markets to the New Jersey equity markets was through  fiber optic lines that allowed equity prices to respond within 7.25-7.95 ms of a price change in Chicago~\cite{Adler}. In August of 2010, Spread Networks 
introduced a new fiber optic line that was shorter than the pre-existing routes and used lower latency equipment. This technology reduced Chicago-New Jersey latency to approximately 6.65 ms~\cite{Steiner,Adler}.

Beginning in approximately March of 2011, the data indicate a further progressive ($\sim 2\, {\rm ms}$) decline in inter-market response latency that we attribute to two factors.  The first factor is the deployment of custom-built microwave communications networks which take advantage of the greater speed of light through air than through glass, and can reduce the inter-market latency by up to approximately 2.5 ms relative to the best available fiber technology. Using FCC records and other data, we document the licensing of at least 15 custom microwave networks between Chicago and the New York area and estimate their latencies to fall in the 4.2-5.2 ms range. Our analysis indicates, however, that most of these links are far from optimally designed, taking far more hops and less direct routes than necessary given available radio hardware and existing towers; this sub-optimal design adds both latency and cost. With respect to the better designed links, we expect that additional improvements in microwave routing and technology to reduce inter-market latency to approximately 4.03 ms, or 0.1 ms above the minimal 3.93 ms speed-of-light latency. At that point, further latency improvement may become prohibitively expensive and difficult, and factors other than long-distance latency, such as ``last mile'' costs and latencies, bandwidth, and link reliability, may begin to dominate over the speed at which information flows between the two markets. 
 
A second factor is the possible evolution of predictive algorithms that statistically anticipate Chicago futures price changes before that information can reach the equity market engines. We observe a signal consistent with the emergence of such algorithms by documenting correlations that occur over timescales {\it shorter} than the theoretical limit of 3.93 ms for light to travel between the Chicago futures market and the New Jersey data centers. Alternately, firms that trade simultaneously in geographically separated markets may coordinate the issuance their orders with respect to the GPS time frame. Such coordinated activity, if it occurs, would make it appear as if information is traveling between Chicago and New Jersey at faster than the speed of light. 

We are also able to estimate the cost of the investments made in order to generate these increased speeds. Spread NetworksÕ fiber optic link is estimated in media reports to have cost approximately \$300 million~\cite{Steiner}, not including the additional cost of ``lighting" dark fibers.
FCC licensing data, frequency prior-coordination notices, and industry sources suggest that of order 20 entities are actively constructing or have already launched microwave links connecting the Chicago and New Jersey areas; notably, the 15 fully-licensed paths we have reconstructed constitute of order 1/20th of all microwave miles licensed over the past two years in the US. At an estimated average capital cost of \$8 million per link, the HFT sector appears ready to spend an additional aggregate of \$160 million to reduce inter-market latency from 6.65 ms to approximately 4.1 ms, at best. We further estimate that a relatively small additional expenditure of $\sim$\$5 million could allow the better designed current microwave networks to come within 0.1 ms of the speed-of-light limit on the Chicago-New Jersey route. 

The remainder of this paper is devoted to providing an in-depth treatment of the foregoing issues. In \S II, we describe our analysis methods for financial tick data and show how they can be used to provide inter-market latency estimates. In \S III, we provide a detailed assessment of the development and the capabilities of the microwave networks that have recently been licensed between New York and Chicago. In \S IV, we combine the {\it internal} latency measurements from the tick data with the {\it external} measurements from the network analyses to show how the Chicago to New York latency has evolved during the past 2.5 years. In \S V, we estimate overall costs that have accompanied the latency improvement, and we discuss the prospects for further development in this fast-moving sub-segment of the financial industry.

\section{Latency Measurement}
\label{sec-measure}

Long-distance telecommunication of financial data is relevant to equities, futures, options, foreign exchange, news, and potentially other market data.  We focus in this paper on correlations between futures trading in Chicago and equity trading and order book response in the New York area exchanges.  These correlations will give a relatively direct measurement of the speed with which trading algorithms in New York are acting in response to events in Chicago.

Our analysis is based on three sources of tick data. The first source is market depth data for the E-Mini S\&P 500 Futures contract purchased from the Chicago Mercantile Exchange (CME)\footnote{http://www.cmegroup.com/market-data/datamine-historical-data/marketdepth.html}. These data are recorded and timestamped at the Globex matching engine, currently located in Aurora, Illinois (longitude -88.24$^{\circ}$ W, latitude 41.80$^{\circ}$ N). Session data is written to ASCII files using the FIX format specification, and then compressed. Level-2 order book activity to a price depth of 5 is captured, along with trade records and other information relevant to recreating the trading session. All order book events are time stamped to millisecond precision time signals propagated from GPS receivers. The GPS system itself provides a globally synchronized timeclock, accurate to of order 50 nanoseconds, tied to a reference frame that rotates with the Earth and properly accounts for both special and general relativistic time dilation; its accuracy is easily sufficient for our purposes as long as the GPS time is correctly propagated to the matching engine.

It is commonly stated that price formation in the United States equity markets occurs in the near-month E-Mini S\&P 500 futures contract, which is traded at the CME and valued at $ \$50\times$~the numerical value of the S\&P 500 Stock Index on the contract expiration date. The E-mini contract  trades on the March quarterly cycle (March, June, September, and December) and expires on the third Friday of the contract month. On the so-called ``roll date'', eight days prior to expiry, both liquidity and price formation shift to the contract with the next-closest expiry date. Several million E-mini contracts are traded each day, corresponding to dollar volumes that frequently exceed \$200 billion.

Our second source of tick data is the NASDAQ TotalView-ITCH historical data feed recorded at the Nasdaq-OMX matching engine currently located in Carteret, New Jersey (longitude -74.25$^{\circ}$ W, latitude 40.58$^{\circ}$ N). These data are composed of a series of binary number-format messages employing the ITCH 4.1 specification and encompass  tick-by-tick information for all displayable orders in the NASDAQ execution system (see~\cite{nasdaq}). Messages are time-stamped to nanosecond precision. 

\begin{figure*}[ht]
 \centering
 \includegraphics[height=11.5cm,angle=0]{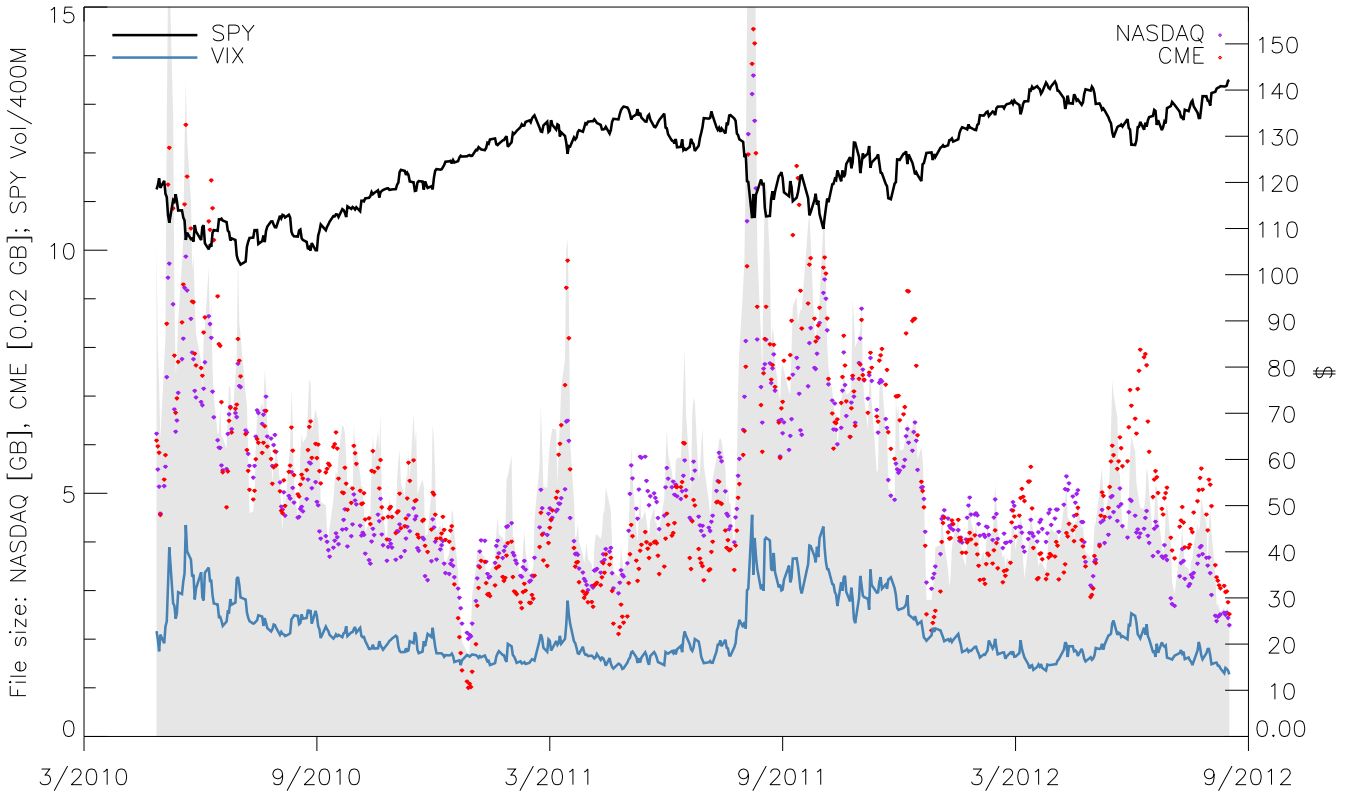}
\caption{ \footnotesize{Compressed file sizes for NASDAQ and CME data during the April 27th, 2010 to August 17th, 2012 period covered by the analysis are shown as purple and red dots, respectively. The data set comprises 558 individual trading days. Note that NASDAQ files are consistently $\sim \, 50 \, \times$ larger than the FIX files supplied by the CME. For comparison, we also chart the SPY share price, the value of the VIX index, both based on daily closing values, and the traded daily volume in SPY. File sizes are seen to be a generally good proxy for the VIX index.}}
\label{fig-filesize}
\end{figure*}

The TotalView-ITCH historical data average 4.13${\rm GB}\,{\rm day}^{-1}$ after gzip compression. The total size of the NASDAQ data used for this study is 2.33TB compressed, and $\sim$20TB uncompressed. In Figure~\ref{fig-filesize}, we plot the compressed file sizes for the full complement of Nasdaq data (in GB)  and the CME data (in units of 0.02 GB to facilitate comparison). Over the period covered by our analysis, the Nasdaq and CME messaging volumes correlate very well, with the Nasdaq data rate consistently running $\sim50\times$ higher than that from the CME. The initial spike in the time series charted in Figure~\ref{fig-filesize} is associated with volatility surrounding the so-called ``Flash Crash" of May 6th, 2010. A second major peak corresponds to the market volatility experienced in August 2011.

A third, higher-level, source of tick data has been obtained from the NYSE Technologies Market Data Custom Trade and Quote Services\footnote{http://www.nyxdata.com/Data-Products/Custom-TAQ/}, and consists of trades in the symbol SPY aggregated across the Consolidated Market System, with price and trade size information time-stamped to the millisecond. 
Although the NYSE data presents a more coarse-grained, lower-resolution view than the Nasdaq feed, it permits measurement of the broader equity market response to activity at the CME, and it is also useful for cross-checking analyses that employ the Nasdaq historical feed.

Our data from the CME, Nasdaq, and NYSE cover 558 trading days that occurred between April 27th, 2010 and August 17th, 2012, inclusive.

Changes in the traded price at the CME have a material effect on the pricing of stocks at the New Jersey exchanges. By correlating order book activity in the equity markets with traded E-mini upticks and downticks, we can estimate the speed of information propagation between the exchanges, and chart its evolution over time. We adopt the following procedures, which employ the Nasdaq, NYSE, and CME tick data described above.

We first step through the CME trade and quote data for a given day. At the end of each millisecond during the period (9:30 AM ET through 4:00 PM ET) when both the CME and the equity exchanges are trading, we screen for the occurrence of near-month E-mini futures trades in which the most recent traded price at the end of a millisecond interval (which we refer to as the ``in-force" trade for a given millisecond) exhibits an increase in price over the most recent in-force trade from a previous millisecond. On typical days during the period of our analysis, such end-of-millisecond based price-increasing events occur about 10,000 times per day (or in about about one out of every 2,000 individual millisecond intervals). In a situation where the E-mini's traded price experiences rapid fluctuations within a millisecond interval, but the in-force trade registers no net change, then our screen registers no response.

When a millisecond interval that ends with a price-increasing in-force CME trade has been identified, our algorithm examines the Nasdaq historical feed for correlated activity associated with a specified equity symbol. For example, the SPY instrument (State Street Advisors S\&P 500 ETF) is a useful choice because it has very high liquidity, and because by design, it closely tracks the S\&P 500 index. Alternately, any other correlated or anti-correlated equity can be used. In each of the 30 millisecond-long intervals prior to, and in each of the 30 millisecond-long intervals following the CME price-increasing trade, we calculate the net number of shares that have been added to the bid side of the SPY limit order book at the three price levels corresponding to (i) the last Nasdaq exchange-traded SPY price, (ii) Êthe last Nasdaq exchange-traded SPY price + \$0.01 , and (iii) the last Nasdaq-traded price - \$0.01. In addition, in each of the same sixty millisecond-long bins surrounding the CME event, we also calculate the net number of shares that have been removed from the three levels of the ask side of the SPY limit order book at prices corresponding to the last Nasdaq-traded SPY price and that price $\pm$~\$0.01. We then add $\delta_l=\,$(added+removed)~to an array that maintains cumulative sums of these deltas as a function of lag (from -30~ms~to~+30 ms). 

The foregoing procedure is also followed for price-{\it decreasing} in-force trades observed in the near-month E-mini contract. In the case of these declines, however, we add $-1\times\delta_l$ to the array that maintains the cumulative sums. This facilitates the combination of both price increases and price decreases into a single estimator.

The resulting composite array of summed $\delta$'s, with each element divided by the total number of price-changing E-mini trades, corresponds to what we will denote the measured ``liquidity signal" for a given day. An error estimate at each lag for a given day is calculated using bootstrap resampling (e.g.~\cite{numrec}) over the observed price-changing events. Although we generally consider one-day intervals, finer or coarser-grained sampling can be employed.

\begin{figure*}[ht]
 \centering
 \includegraphics[height=8cm,angle=0]{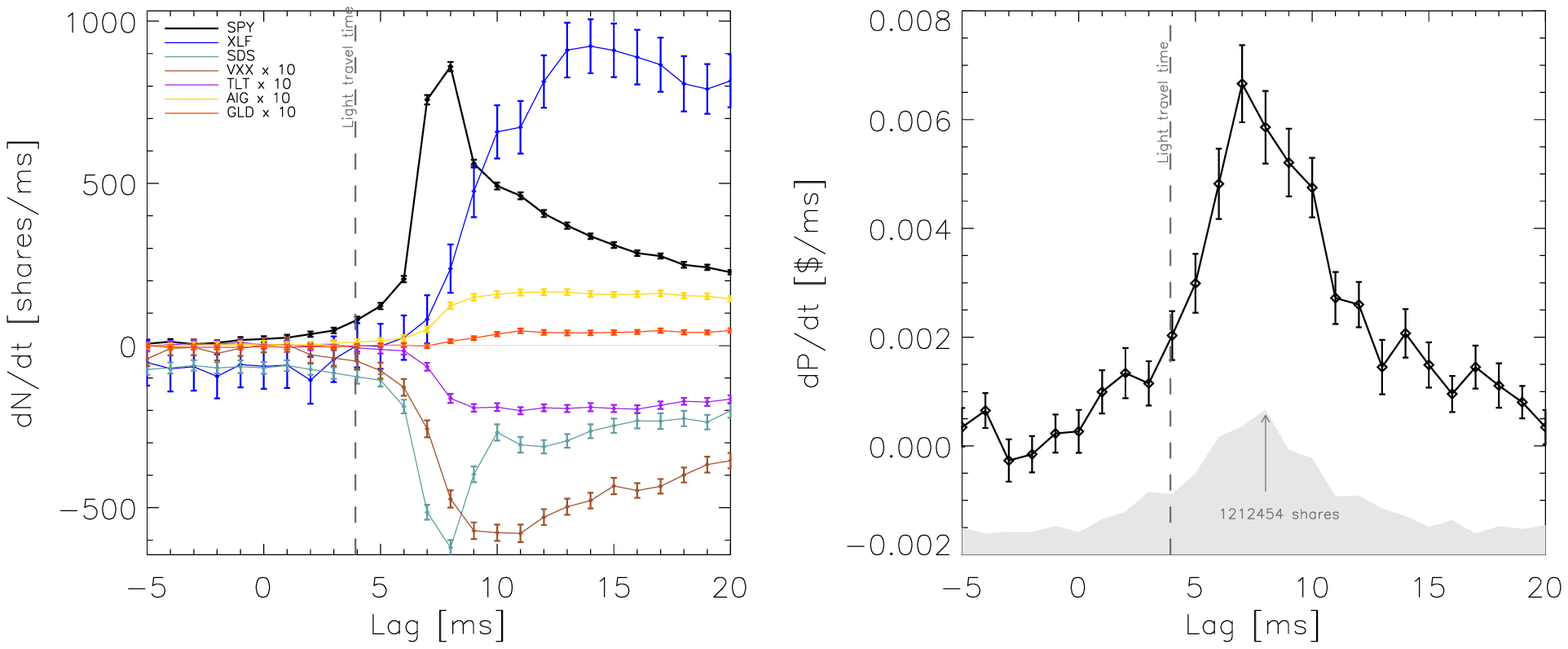}
\caption{ \footnotesize{{\em Left:} Liquidity response evaluated for February 15th, 2012 at the Nasdaq matching engine (located in Carteret, New Jersey) to price-changing trades in the near-month E-Mini future executed at the Globex platform in Aurora, Illinois. Full-day liquidity responses (see text for methodology) are shown for SPY, XLF, and SDS, along with VXX, TLT, AIG, and GLD (with measured values for the latter four securities multiplied by 10 in order to clarify the structure of the response). The two trading venues are separated by $d=1,179~$km. The speed-of-light travel time between the two locations is $t=c/d=3.93~{\rm ms}$. {\em Right:} Trade response (see text for methodology) for SPY as measured from Consolidated Market System data to price-changing E-Mini trades for February 15th, 2012.
}}
\label{fig-oneday}
\end{figure*}

The solid, black curve in the left panel of Fig.~\ref{fig-oneday} shows the result of following the above-described correlation procedure to compute liquidity responses for trading that occurred on Feb 25th, 2012. This example trading day was characterized by market volume that was typical of daily volume averaged over the full 2010-2012 period covered by our analysis. Fig.~\ref{fig-oneday} indicates for example, that on average, during the 7th millisecond following a price-changing trade at CME, there was a $\sim$800-share liquidity response at the inner levels of the SPY limit order book on the Nasdaq exchange. That is, on average, a price increase at CME was met by high-frequency traders removing $\sim400$ shares of SPY from the ask side of the order book and adding $\sim400$ shares to the the bid side of the limit order book. 

Fig.~\ref{fig-oneday} also shows liquidity responses for six additional high-volume instruments traded on the Nasdaq exchange.  Three categories are represented. First, there are two equities with strong inverse correlation, that is, names with large negative $\beta$'s in the language of portfolio theory. These include SDS ($\beta_{\rm SDS}=-1.78$)\footnote{All reported values for $\beta$ were taken from Google Finance, accessed Oct. 1, 2012, and are intended here only for purposes of rough comparison.}, the ProShares UltraShort S\&P500 which uses derivative positions to seek twice the inverse (-2x) the daily performance of the S\&P 500 index, and  VXX ($\beta_{\rm VXX}=-2.69$), the iPath S\&P 500 VIX Short Term Futures ETN, which provides exposure to a daily rolling long position in the first and second month CME VIX futures contracts. Second, are two equities -- TLT (iShares Barclays 20+ Year Treasury Bond ETF, $\beta_{\rm TLT}=-0.29$ and GLD (SPDR Gold Trust ETF $\beta_{\rm GLD}=-0.13$) -- chosen for their $\beta$ values near zero, and hence expectation for low correlation to short-term price fluctuations in the E-mini. The third category consists of AIG (American International Group common shares $\beta_{\rm AIG}=3.43$) and XLF (Financial Select Sector SPDR ETF $\beta_{\rm XLF}=1.49$). These names were chosen for a combination of high volume and high $\beta$.

The liquidity response functions charted in Figure~\ref{fig-oneday} display a number of interesting features. Most importantly, the diagram illustrates that on average, there is a highly significant liquidity response in the equity order books to price-changing trades for the near-month E-Mini contract at the CME. The response for an individual symbol can, in general, have two contributions. The first, and apparently dominant, contribution is the arrival of specific news from the CME that a price-changing E-mini trade has occurred. The speed of this information transfer cannot exceed the speed of light, $c$, which sets a hard $t_{min}=3.93$~ms minimum information propagation time between Aurora, IL and Carteret, NJ. A second contribution to the observed response signal will arise if market participants in Carteret have a statistically significant ability to predict the occurrence of price-changing trades at the CME, or if individual participants place pre-coordinated orders in the physically separated markets. A consistent pattern of successful predictions (or pre-coordinated activity) will generate measurable liquidity shifts prior to the arrival of the actual news. As discernible in Figure~\ref{fig-oneday} and discussed further below, there is unambiguous evidence for such a contribution to the SPY response curve.  Furthermore, it is clear that the shapes of the response curves have a significant dependence on $\beta$. The symbols SDS and SPY, which are most closely tied to the E-mini future, react with large amplitude and short response times. Somewhat surprisingly, the largest {\it integrated} response among the six symbols is observed for XLF. Although this equity is slower to respond than SDS and SPY, its total integrated response is both large and sustained, suggesting that (as of February 15, 2012) the XLF ETF provided important (and perhaps not fully exploited) opportunities for high-frequency electronic market-making strategies.

The right panel of Figure~\ref{fig-oneday} shows a very similar analysis except that rather than computing the change in liquidity in response to price-changing E-Mini trades, we compute the response for a selected equity (in this case, SPY) in the form of price-changing trades (provided via NYSE TAQ data for the US Consolidated Market System covering NASDAQ, NYSE Arca, NYSE, BATS BZX, BATS BYX, EDGX, Nasdaq, CBSX, NSX, and CHX). Specifically, when a millisecond is identified in which the in-force E-mini trade displays a price change from the most recent in-force E-mini trade at the close of a previous millisecond, we look for near-coincident trades in a particular symbol (here, SPY) on the consolidated market system tape. In each of the 30 millisecond intervals prior and following the CME event, if a price-changing trade in the selected symbol has occurred, we add the observed change in the traded price ($\delta_t$) to an array that maintains a cumulative sum of these deltas from ${\rm lag}=30~{\rm ms}$ to ${\rm lag}=+30~{\rm ms}$. The resulting {\it trade-trade} responses have larger error bars because most orders submitted to the matching engine (generating liquidity response) do not actually result in fills (that generate a trade response). 

\begin{figure*}[ht]
 \centering
 \includegraphics[height=3.7cm,angle=0]{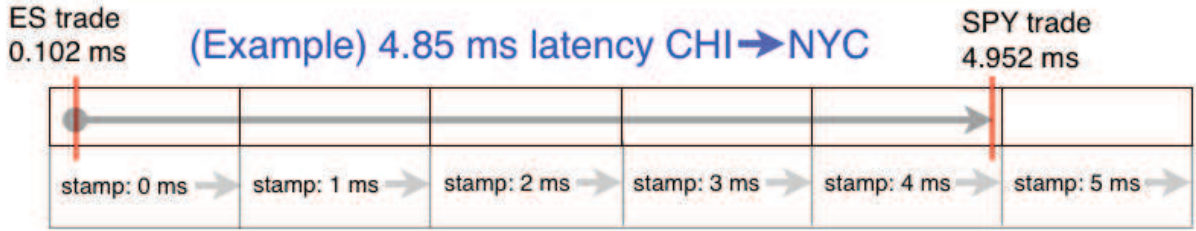}
\caption{ \footnotesize{An end-to-end propagation time between $N$ and $N+1$ ms is reflected in contributions to both bin $N$ and to bin $N+1$. Here, for purposes of example, a hypothetical line travel time, $t_{1}=$4.85ms produces contributions to bins $N$ and $N+1$, with 85\% of the signal falling in the 5ms bin.}}
\label{fig-rounding}
\end{figure*}

An important note regarding all of our response curves is illustrated in Figure~\ref{fig-rounding}, which indicates that on average, given our methodology, and given the millisecond resolution of the order book time stamps, an end-to-end propagation time between $N$ and $N+1$ ms is reflected in contributions to both bin $N$ and to bin $N+1$. In the analysis of this paper, we do not attempt to gain information with sub-millisecond precision by tracking the ordering among all of the messages that share the same millisecond-resolution time stamp. 

\begin{figure*}[ht]
 \centering
 \includegraphics[width=18cm,angle=0]{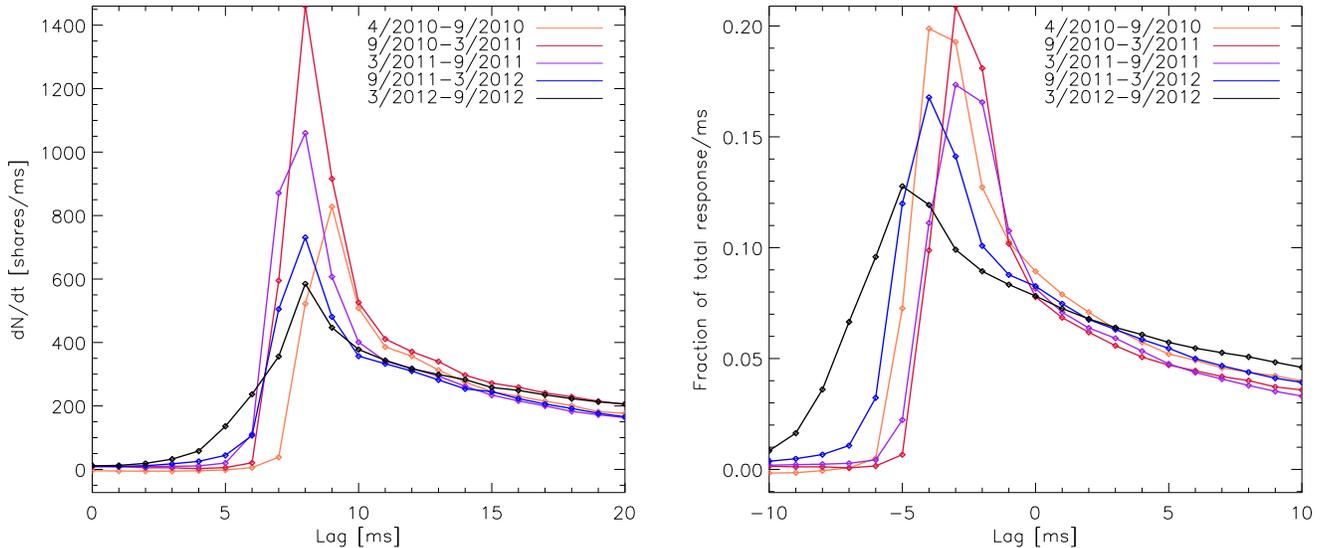}
\caption{ \footnotesize{SPY liquidity response measured at the Nasdaq to price-changing trades in the near-month E-Mini future. {\em Left:} Each point on each curve represents the median of all data points for a given lag during a particular time period; the curves thus represent averaged response curves over ~6 month periods. {\em Right:} The same response curves, but each curve has been normalized so that it sums to unity for lags of 0-20 ms, and also shifted so that its 50\% response time is at Lag=0; these curves thus illustrate the change in {\em shape} of the response, independent of scaling or time lag.}}
\label{fig-trade_liq_resp_curves}
\end{figure*}

\begin{figure*}[ht]
 \centering
 \includegraphics[width=18cm,angle=0]{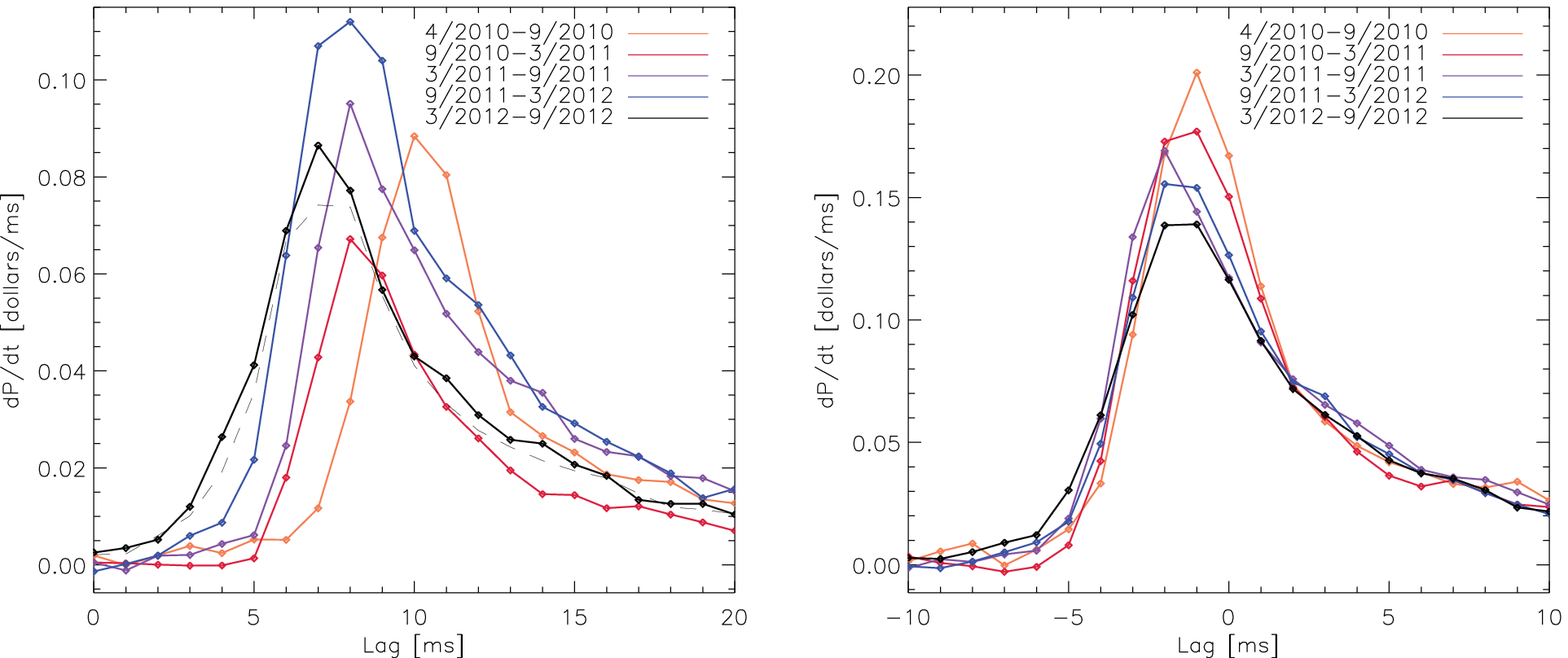}
\caption{ \footnotesize{As in Fig.~\ref{fig-trade_liq_resp_curves} but for the traded price response to the CME price-changing trades. In addition, the dashed line shows the same analysis for the most recent epoch, using trades at the NASDAQ exchange only, as recorded in the ITCH4.1 data.}}
\label{fig-trade_trade_resp_curves}
\end{figure*}

The statistical errors on several of the curves in Figure~\ref{fig-oneday} are quite small, but even in these cases the day-to-day variation significantly exceeds the error estimate for the signal at each lag, so it is useful to accumulate data over a number of days. The left panels of Figures~\ref{fig-trade_liq_resp_curves} and~\ref{fig-trade_trade_resp_curves} show the ``liquidity" and ``trade" response for SPY accumulated over 6-month intervals, each containing approximately 125 trading days. These curves show a significant changes over the past several years; to analyze these changes it is useful to separate them into the curves' relative amplitudes, shapes, and positions along the ``lag" axis.  

The response amplitude can be quantified by integrating the total measured response for an equity symbol (as plotted, for example in Figure~\ref{fig-oneday}) from $t_0$=0 ms through $t_{f}=30$ ms (or some other fixed point that is substantially later than the bulk of the equity market response to a CME price-changing trade), to give the total 
$$
T= \int_{0}^{t_f} {dP\over{dt}}dt\, .
$$ 
To examine shifts in the lag (which bear most directly upon latency in communication), we can calculate the time required to cover a specified fraction, $X$ of the total response. For trade-trade response, we calculate the latency, $t_{t\,X}$ such that
$$
{1\over{T}}\int_{0}^{t_{t\,X}} {dP\over{dt}}dt =X \, ,
$$
with an analogous estimate for $t_{l\,X}$ estimated from the trade-liquidity response.  In the above equation, linear interpolation is used to numerically integrate over the discretely sampled millisecond bins.

The right panels of Figures~\ref{fig-trade_liq_resp_curves} and~\ref{fig-trade_trade_resp_curves} panels show aggregated responses as in the left panels, but in which for each day, the response curve has been divided by the total $T$, and shifted in time by the time $t_{l50}$ at which 50\% of the response has accumulated. This allows a clear comparison of the {\em shape} of the response across the various epochs.

These figures exhibit a number of interesting features.  Both trade and liquidity responses are composed of a rapid rise over $\lesssim 5\,$ms, followed by a much more gradual falloff over tens of ms (though the liquidity response's rise is generally faster and its falloff slower.)  The curves have evolved fairly differently, however, over the past several years, with the trade signal retaining a nearly-constant shape that has steadily shifted to lower latency, while the liquidity signal has significantly broadened both before and after its peak.  

In interpreting these correlations we will make a few core assumptions.  The first is that the correlation arises from the causal influence of CME price-changing trades on the NY equity markets.  In principle, as mentioned above, a correlation can be created by a common cause (such as orders being transmitted simultaneously or with some time offset to both Chicago and NY markets), though this would be an unlikely interpretation of the bulk of the signal. The second is that the timestamps included in the data accurately represent the time at which a given order was placed or trade executed, as provided by the globally synchronized clock of the GPS system; this is warranted by the exchanges but difficult for us to verify directly.\footnote{Inaccurate time-stamping can, of course, bias the inferred speed of signal transmission; for an unfortunate and high-profile recent example see, e.g.~\cite{Cartlidge}. 
}  All exchanges send fill messages to participants before they report trades to the market data tape. Depending on the load at the matching engine and other factors, there is a potentially significant and stochastic delay between the moment when a trade is time stamped, and when it is published at the exchange gateway. It is therefore distinctly possible that a sophisticated participant who receives a fill message from the CME can bypass significant latency, drift, and jitter, and send the information directly to the equity exchanges. If this is the case, then the fact that only {\it one} participant receives advance notice of a fill will act to suppress the magnitude of the early response at the equity exchanges in New Jersey, and, further, would have very interesting consequences for the latency competition between microwave networks. 

Comparison of timestamps for identifiable trades recorded by both the NASDAQ ITCH4.1 system and the ARCA/CMS system indicate sub-ms overall offset, but with 0-2 ms jitter in the relative time recorded; we will attribute this to relative delay in processing orders between one system and the other, as the exchanges included in the CMS are geographically separated by up to $d\approx 50\,$km ($\gtrsim 250\mu\, s$ by fiber). The response plotted may thus represent the convolution of a ``true" response with a distribution with width of order $1\,$ms; however, assuming GPS accuracy of timestamps, this jitter should never allow a New York response to a Chicago event to be recorded prior to the actual propagation time.  Indeed, as shown in Figure~\ref{fig-trade_trade_resp_curves}, the trade response using the ITCH4.1 system (dashed lines) is quite consistent with that of the ARCA/CMS system in the most recent epoch. (Similar curves for other periods are also broadly similar, though sub-ms lateral offsets in both directions appear for some epochs; these may reflect actual changes in trading amongst different exchanges, but we do not analyze this in detail here.)

Under these assumptions, we can see that both signals show no significant evidence for information propagation faster than 7\,ms in the earliest dates, but have seen a downward shift in response time of $\sim 3$ ms since. This shift can be attributed to two potential causes.  

First, decreases in actual communications latency, will drive the signal toward the light-travel time of 3.93\,ms, which is the fundamental limit to a direct causal signal between the CME and the NJ matching engines.\footnote{It is amusing to note that even a signal beamed directly through the Earth (see, e.g.~\cite{neutrinos}) could cut only $\sim 1\,km$, or $\sim 3\mu s$ from this latency.} 

\begin{figure}[ht]
 \centering
 \includegraphics[width=9cm,angle=0]{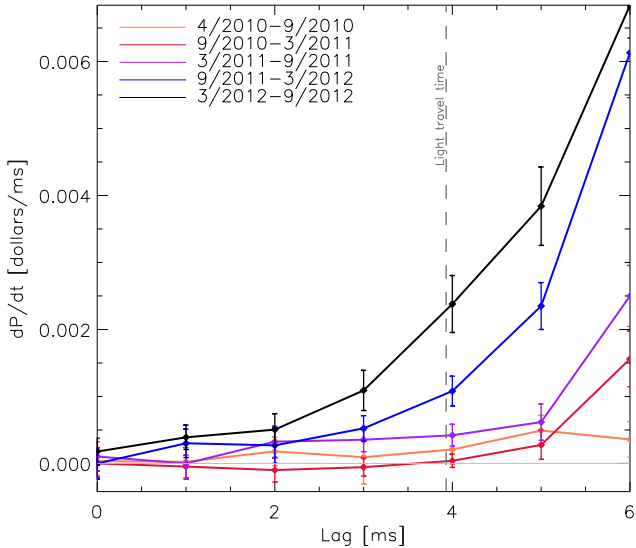}
\caption{ \footnotesize{As in Fig.~\ref{fig-trade_trade_resp_curves} but showing only Lag=0-6\,ms, and with 95\% confidence intervals shown.  Of note, there is a highly statistically significant rise prior to 4\,ms appearing in late 2011/early 2012, and growing subsequently. A similar signal can be seen in the liquidity response curves (not shown.)
}}
\label{fig-presig}
\end{figure}

Second, if a trading algorithm can predict (on the basis, for example, of changes to the balance of contracts on bid and offer in the E-Mini order book) when a price-changing trade will occur, and trade on the basis of that prediction, this will appear as a contribution to the signal that can occur prior to the 3.93\,ms limit.  Such a signal is, in fact, evident in both the trade and liquidity signals starting in 2012. Figure~\ref{fig-presig} shows the early part of the trade response with 95\% confidence intervals plotted as error bars; these are based on bootstrap resampling of the daily data points aggregated into each lag bin.  

While it is difficult to disentangle improvements in prediction (or, alternately, an increase in order coordination) from improvements in communication latency, the evolution of these curves strongly suggests both processes are taking place. As described in the next section, the high value of short response time to information from the CME has led to significant investment in faster communication channels via microwave.

\section{Microwave Networks}
\label{sec-mwnetworks}

The speed of electromagnetic information transmission through air, including microwave (MW) at $\sim 10^3-10^5\,$MHz, occurs a factor of $1/n$ faster than in fiber of refractive index $n$.  Long-distance MW networks can moreover be significantly straighter (closer to the geodesic path between the two endpoints) than fiber networks because MW paths can also travel over small bodies of water and other regions in which it is difficult to lay fiber. Thus MW networks 
can in principle offer significantly lower latencies than fiber on overland routes.  

To be successful, however, such networks must overcome several technical obstacles:
\begin{itemize}
\item Commercially available digital MW radios are a mature and highly efficient technology (operating close to the Shannon limit), but have generally been optimized primarily for reliability, throughput, and cost, but {\em not} latency.  Nearly all MW radios on the market as of 2010 therefore added a latency $L_{\rm rad}$ of many tens or even thousands of microseconds of latency per MW hop.
\item Private MW networks in the US are limited to particular channels, predominantly 30 or 40 MHz channels in the 5925-6875 MHz and 10700-11700 MHz bands.\footnote{Higher-frequency channels are available, but susceptible to significant rain outage; at lower frequencies it is difficult to obtain a large enough frequency channel free of interference to sustain high data rates.}  Use of this limited frequency space is governed by an FCC licensing process that constrains the bit rate, power, modulation, and other characteristics of MW data channels; obtaining the requisite licenses can be difficult in congested regions (as described below, the NY-Chicago corridor as a whole has become such a region.)
\item MW networks are susceptible to outage due to propagation effects such as multi-path interference, rain fades, and significant variations in the atmospheric refractivity gradient, as well as equipment and power failure, and physical damage to radios or MW towers.  Commercial networks can nonetheless achieve very high (99.999+\%) uptime; however, this entails a conservative and redundant design strategy  generally requires a tradeoff with both bit rate and latency.
\end{itemize}
The next section describes the basic physics and engineering considerations for long-distance low-latency MW networks; we then discuss existing and planned financial data networks in light of these considerations.

\subsection{Design considerations for latency-optimized MW networks.}

Consider an $N-$hop MW network of total length $D_{\rm tot}$ between endpoints $A$ and $B$ running at frequency $f$.  The achievable hop length (which determines $N$) is limited primarily by Earth's curvature, and secondarily by reliability considerations.  The curvature constraint arises because the microwave path must clear the Earth as well as obstructions of height $h_{\rm obs}$ by approximately one Fresnel-zone width $h_{\rm fres}$, and the Earth's curved surface can be treated as a `bulge' of height $h_{\rm Earth}$ that a straight-line path must clear.  At the midpoint of a hop of length $D$, we have
$$
h_{\rm fres} \simeq 8.7\,m\left(D\over 1\,{\rm km}\right)^{1/2}\left({f\over 1\,{\rm GHz}}\right)^{-1/2}
$$
and
$$
h_{\rm Earth} \simeq {1\over 50}\,m\left({D\over 1\,{\rm km}}\right)^2K^{-1},
$$
where $K$ accounts for the gradient of the atmospheric refractive index, which effectively multiplies the Earth's radius by a factor $K$.  (For this and other standard equations in microwave engineering, see e.g., Ref.~\cite{Manning}.)  Assuming tower of height $h_{\rm tow} \lesssim 100\,m$, $f \approx 6\,$GHz, $K < 4/3$, and $h_{\rm obs} \gtrsim 10\,m$, the constraint that $h_{\rm tow} > h_{\rm fres} + h_{\rm Earth} + h_{\rm obs}$ requires $D \lesssim 70\,$km over flat terrain.  (Intervening terrain peaks and valleys can increase or decrease this number, and it is also limited by tower availability, but this is a fairly accurate estimate, as well-planned networks spanning the $\sim 1200$\,km between New York and Chicago data centers require $\gtrsim 20$ hops.)

For a MW network to provide a significant advantage over fiber of refractive index $n$ along the same path requires that the equipment latency $L_{\rm rad}$ satisfy $D/c + L_{\rm rad} \ll n D/c$, or $L \ll 100\,\mu s$ if $n \approx 1.4$.  Although not common until recently, this is easily achievable.  Microwave radios utilizing 30-40\,MHz channels and 64-256QAM modulation run at a bit rate of $\sim 140-190$ Mbps.  Digital radio latency is dominated by (a) forward error correction (FEC) that buffers typically 32-256 bytes of data, (b) interleaving data to increase the effectiveness of this FEC for bursty errors, and (c) ethernet switches that buffer $\sim 64-1024$\,byte ethernet packets.  Buffering and reading out $B$ bytes at bit rate $\dot B$ requires
$$
L = 10.24\, (B/64\,{\rm bytes}) (\dot B/ 100\,{\rm Mbps})^{-1}\,\mu s.
$$
Thus, once the interleaver is removed, radios with FEC and ethernet packets can achieve $L \sim$(tens) of $\mu$s.  Radios optimized for latency with these removed are advertised at $L < 10\,\mu$s, and there is no barrier in principle to radios with  $L \lesssim 1\,\mu$s. (Moreover, analog microwave repeaters exist and can amplify and retransmit a data stream with latency $L \ll 1\,\mu$s.)

Along with the radio latency and hop number, a MW path's overall latency is determined by the path length.  As the Earth is spherical to good approximation, the shortest distance geodesic path along its surface is approximately a geodesic of length $D_{\rm geo}$.  The `excess distance' $D_{\rm ex}$ that a path requires relative to this geodesic distance is determined by terrain (such as bodies of water), availability of towers, and skill in constructing a string of rentable (or buildable) towers with proper clearances and available frequencies.  Contributions to $D_{\rm ex}$ arise due to the change in heading $\delta\alpha$ between two successive MW hops.  By simple geometry, for a two-hop network\footnote{A three-hop network with three equal hops and two successive azimuth changes of the same magnitude but opposite sign, the result is $\delta D_{\rm ex}/D \simeq (\delta\alpha)^2/4$, and a general network with azimuth changes $\approx \alpha$ should generally lie in this range.} with $\delta\alpha \ll 1$ and hop length $D \ll R_{\rm Earth}$, the contribution $\delta D_{\rm ex}$ is given by $\delta D_{\rm ex}/D \simeq (\delta\alpha)^2/8.$ Relatedly, in a hypothetic two-hop route the perpendicular excursion from the geodesic is of order $\sim \sqrt{D_{\rm tot}D_{\rm ex}/2}$.  
(This indicates that, for example, a 1200 km route can deviate by $\approx 70\,$km north or south of the geodesic while maintaining $D_{\rm ex} \lesssim 10\,$km.)  Thus  there is some latitude in the gross path of a low-latency MW route, but care must be taken to minimize the relative azimuths  between successive hops. 

To summarize, a  MW route carries an `excess latency' $L_{\rm ex}$ beyond the theoretically possible overland latency of $\simeq D_{\rm geo}/C$ of
$$
L_{\rm ex} = N L_{\rm rad} + D_{\rm ex}/c,
$$
where from the considerations above, and by comparison with deployed routes (see below), $N \sim D_{\rm tot}/(30-60\,{\rm km})$, $L_{\rm rad} \sim 1-20\,\mu$s, and $D_{\rm ex} \sim (0.001-0.01)D_{\rm tot}$.  This leads to a range of 
$$
L_{\rm ex} \approx 60-1200\,\mu s
$$
along the NY-Chicago route.  For reference, the currently-fastest fiber route requires $> 6.55\,$ms,\footnote{This latency is quoted from 350 E. Cermak Road (longitude -87.62$^{\circ}$ W, latitude 41.85$^{\circ}$ N)  in Chicago to data centers in NJ; the current CME data center in Aurora IL is $\sim 60\,$km more distant.} or a latency excess of $L_{\rm ex} \gtrsim 2700\,\mu$s between the CME and Nasdaq matching engines.

\subsection{Existing and planned Chicago-NY MW networks}

Although firms constructing MW networks for financial data transmission generally do not broadcast the details of their activities, all operating MW links in the FCC licensed bands must legally disclose the details of their links to the FCC via its licensing system, and that license information publicly available.  Using software tools we have developed to parse this database, we have been able to reconstruct the routes of 15 full MW routes with licenses either granted or applied for by September 1, 2012.\footnote{Note that such routes are subject to modification by new or revised licenses, so this is a snapshot of a fluid picture. Moreover, some routes take multiple paths for some segments, and terminate at several different data centers; where possible we have assumed the shortest path for our estimates.}  

\begin{figure*}[ht]
 \centering
\includegraphics[width=6.5in,angle=0]{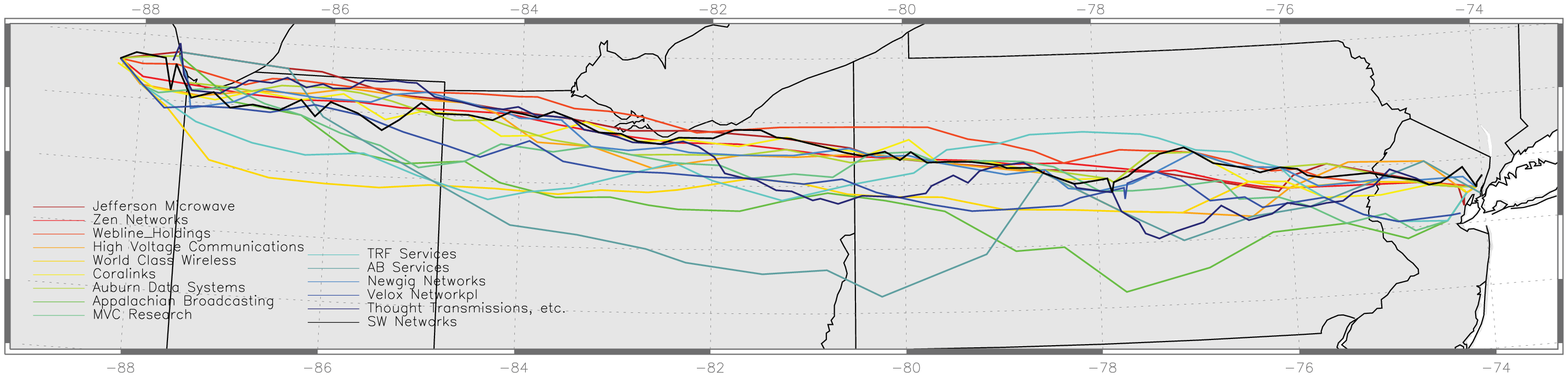}
 \includegraphics[width=6.5in,angle=0]{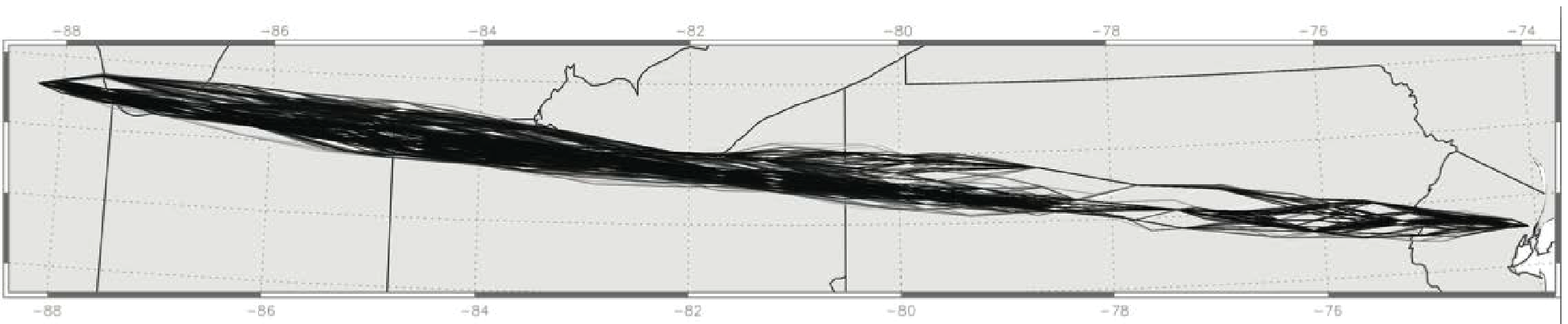}
\label{fig-routes}
\caption{{\em Top:} Fifteen microwave routes between Chicago and NJ data centers with licenses or license applications in the FCC 6 and, 11, and 18 GHz bands as of September 1, 2012. Routes are labeled by the legal entity that submitted the application; order and color is by increasing latency as computed in Table~\ref{tab-routes}. {\em Bottom:} 7,700 example routes planned across existing towers with valid clearances. All of these are shorter than all existing MW links, showing that current routes are generally far from optimal in communicating information at low latency.}
\end{figure*}

	These routes are and listed in Table~\ref{tab-routes} and illustrated in Figure~\ref{fig-routes}.  The table lists the routes in chronological order of the first applied-for license.\footnote{Note that this does not necessarily correspond to the order in which the routes have come into operation -- indeed, a number are almost certainly still under construction.} The ``registered licensee" is generally a company spawned in order to separate the route from the identity of its user/operator, though some  do correspond to publicly disclosed carriers.  Total and route distances $D_{\rm tot}$ and $D_{\rm geo}$ are computed between the endpoints of the MW route, though the information channel may continue by other technologies such as fiber, free-space optical, or other RF channels for short distances beyond these endpoints, to connect to data centers.  We also neglect path length due to cabling between radios at each tower, which adds on average $\approx 150\,m$ and $\approx 0.75\mu m$ per tower (assuming a signal speed of 0.66\,c in the cabling), or $\approx 3-12\,km$ and $\approx 15-60\,\mu s$ over a route. To estimate total latency excess $\Delta L_{\rm ex}$, we have provided figures for a fiducial value of $L_{\rm rad} = 10 \mu$s; other values\footnote{FCC records also list the radios used in a given link; however latency figures for these radios are generally confidential.} can be obtained by scaling the ``equipment latency" column by $L_{\rm rad}/(10\,\mu s)$ and adding to the ``distance latency" $L_D$ column.   The final column lists the maximal licensed bit rate in Mbps. Most routes are licensed for several modulation rates, and some for several frequency channels; we list the total throughput combining all channels at maximum bit rate. All distance figures are in km, and all latencies in $\mu$s.

\begin{table*}\footnotesize
\begin{center}    
    \begin{tabular}{ | l | l| l | l || l | l | l || l | l | l || l |}
    \hline
    Registered Licensee & \# files & First file & N & $D_{\rm geo} $ & $D_{\rm tot} $ & $D_{\rm ex} $ & $L_D$ & $N L_{\rm rad}$  &  $L_{\rm ex}$ & Mbps  \\ \hline\hline
   AB Services & 43 & 9/22/10 & 18 & 1140 & 1269 & 129 & 430 & 180 & 610 & $2\times 155$\\  \hline
   Auburn Data Systems & 98 & 12/11/10 & 40 & 1186 & 1224 & 38 & 127 & 400 & 527 & 164 \\ \hline
   Thought Transmissions\footnote{This route is licensed by three separate entities: ``Thought Transmissions", ``Fundamental Broadcasting" and ``Comprehensive Wireless"} & 213 & 12/23/10 & 81 & 1104 & 1226 & 122 & 406 & 810 & 1216 & 135 \\ \hline
   %84+61+68
   Appalachian Broadcasting & 63 & 6/28/11 & 26 & 1128 & 1211 & 83 & 260 & 277 & 539 & $2\times 191$ \\ \hline
   SW Networks & 219 & 8/2/11 & 77 & 1190 & 1340 & 150 & 500 & 770 & 1270 & 139  \\ \hline
   Webline Holdings & 89 & 9/29/11 & 32 & 1186 & 1200 & 14 & 47 & 320 & 367 & 191 \\ \hline
   Newgig Networks & 73 & 10/5/11 & 37 & 1135 & 1210 & 75 & 250 & 370 & 620 & $2\times 139$ \\ \hline
   World Class Wireless & 386\footnote{This figure includes applications for other routes including Chicago-Washington DC.} & 10/7/11 & 30 & 1184 & 1241 & 60 & 200 & 300 & 500 & 167 \\ \hline
   Jefferson Microwave & 91 & 8/17/11 & 20 & 1187 & 1195 & 9 & 30 & 200 & 230 & $2\times 165$ \\ \hline
   Coralinks & 91 & 11/15/11 & 33 & 1180 & 1236 & 56 & 188 & 330 & 518 & 139  \\ \hline
   High Voltage Commun. & 134 & 2/28/12 & 25 & 1179 & 1232 & 53 & 177 & 250 & 427 & $2\times 167$ \\ \hline
   TRF Services & 77 & 3/7/12 & 31 & 1179 & 1255 & 76 & 253 & 310 & 563 & 167  \\ \hline
   MVC Research & 38 & 3/23/12 & 36 & 1165 & 1224 & 59 & 197 & 360 & 557 & 167 \\ \hline
   Velox Networks & 44 & 3/30/12 & 37 & 1155 & 1247 & 92 & 306 & 370 & 677 & 139 \\ \hline
   Zen Networks & 117 & 11/22/11 & 31 & 1179 & 1188 & 9 & 30 & 310 & 340 & $2\times 191$ \\ \hline
    \end{tabular}
    	\label{tab-routes}	
	\end{center}
	\end{table*}

Several more networks (``China Cat Productions",  ``Pitt Power LLC", ``Converge Towers, LLC") are partially licensed, and several additional networks are in earlier stages of licensing.  In addition, multiple networks exist connecting the Chicago area to Washington, DC, and Washington, DC to the NJ data centers, as can be discerned from Fig.~\ref{fig-maps} below.  

These routes  all show a significant latency advantage relative to even a custom-built fiber route, with one-way latencies generally $< 5\,$ms.  (Note that in terms of the listed quantities,  $L_{\rm min} \equiv D_{\rm geo}/c + L_{\rm ex}$ represents the total one-way latency between endpoints for the given $L_{\rm rad}$.  However,  this is not directly comparable between routes with different endpoints, and  comparison with financial data requires estimates of latency between these endpoints and the data center data servers.) These routes are, however, far lower in data throughput -- even the total combined bit rate of $\simeq 3.4\,$ Gbps for all 15 networks is far below the Tbps carried by fiber routes.

The licensed routes exhibit very significant spread in ``excess" latency, by a factor of five, even under the same assumed latency per radio.  This indicates both that these routes are generally far from optimal given the available tower and frequency resources, and that at present `skill' at minimizing these azimuth differences (and maximizing $D$ for each hop) is the prime determinant of path length along the NY-Chicago route.  This is perhaps unsurprising given that conventional hop-by-hop planning of MW routes is poorly suited to solving a global optimization problem of this sort.  Suppose, for example, that an $N$-hop route  is constructed by iteratively choosing the tower with a live link in closest to the correct direction. If each tower has on average $n_{\rm near}$ live links to nearby towers of distance $\sim D_{\rm tot}/N$, then  each link is typically forced to deviate by an amount $\delta \alpha \sim {\pi/n_{\rm near}}$, resulting in 
$$
D_{\rm ex} \sim {1\over 4}\left({\pi \over n_{\rm near}}\right)^2D_{\rm tot}.
$$
The range of $D_{\rm ex}/D_{\rm tot}$ exhibited by typical networks suggests $n_{\rm near} \sim 5-10$ through the parts of the path dominating $D_{\rm ex}$, and this value agrees with estimates of tower density based on FCC data (see Fig.~\ref{fig-maps} below). 

Far better results can be achieved by planning $M$ links at a time: allowing, for example, a choice between the {\em two} towers nearest the right direction gives on average 50\% greater $\delta\alpha$; but now there are {\em four} chances to line up the azimuths of the first and second hops.  These combinatoric factors add up quickly, so a global optimization of the path length yields far better results.  To examine in detail how optimal a path {\em could} be planned, we have run numerical experiments by identifying `live' links between existing towers in the Chicago-NJ corridor, and found the lowest-latency possible path achievable using the tops of all existing towers; this turns out to be an extremely short $D_{\rm ex} \sim 0.2\,$km. However, realistically not all heights on all towers will be available.\footnote{In addition, frequency space between Chicago and NJ is becoming quite crowded.  While this did not hinder early networks, it is likely to be a key constraint in future ones.} We therefore developed a probability model for the availability of both towers of different types, and heights on towers, calibrated via inquiries to tower companies for a sample of towers.\footnote{Thanks to D. Levine for assistance in this project.}  As exhibited in Fig.~\ref{fig-routes}, even with realistic (and probably conservative) probabilities, paths of $N \approx 19-24$ and $D_{\rm ex} \approx 4-10\,km$ are routinely achievable, yielding far lower latencies than most extant networks (although the best of these, Jefferson Microwave and Zen Networks, at least approach the upper end of the $D_{\rm ex}$ range.)  This, along with the large range in existing routes, indicates that at present the primary determinant of MW path latency is the sophistication (or lack thereof) in route planning, and that convergence toward lower-latency routes can and probably will occur in the future.  In particular, given radios or repeaters of $< 2-3\,\mu s$ average latency, a path with $L_{\rm ex} < 100\,\mu s$ could quite plausibly be constructed, perhaps as an upgrade of an existing path to help with frequency or tower-congested areas.

\section{Correlating information propagation with the financial data}
\label{sec-analyze}

It is of substantial interest to know how the line-time between Chicago and New York available to various trading firms has evolved in response to investments in communication infrastructure. In the previous section, we discussed how {\it external} estimates of the Chicago to New York latency can be made by reconstructing microwave routes from the database of FCC licenses, coupled with estimates of the associated radio equipment, last mile, and exchange-associated latencies. A difficulty with this approach is that the existence of an FCC license does not mean that a particular link is actually functioning, making it hard to know whether a particular route is transmitting data; it is also impossible from such data to determine how many firms are using each such route.

A correlation analysis of the exchange-provided tick data, such as developed in Sec.~\ref{sec-measure}, demonstrates the effects of networks that are actually operating, but there are several factors that make it difficult to determine the precise latency of information transport. In particular, the construction of a single latency statistic that is simultaneously compelling and accurate is affected by several factors, including (1) the diversity of both firms and strategies that are combined in the amalgamated market response; (2) the a-causal ``pre-signal'', which occurs when market participants are consistently able to predict and anticipate the price movements of a given security before they occur, or when they are able to coordinate issuance of orders at different exchanges according to a pre-defined schedule; (3) substantial day-to-day variation in market response to exogenous shocks, which we attribute to the effect of evolving algorithmic trading strategies, and possible non-stationary systematic delays between the time stamping of trades and their appearance on the data feed at the exchange gateway (rather than random noise); and (4) strong evolution of both volume and the overall shape of the response curve over the past several years.

\begin{figure*}[ht]
 \centering
 \includegraphics[width=18cm,angle=0]{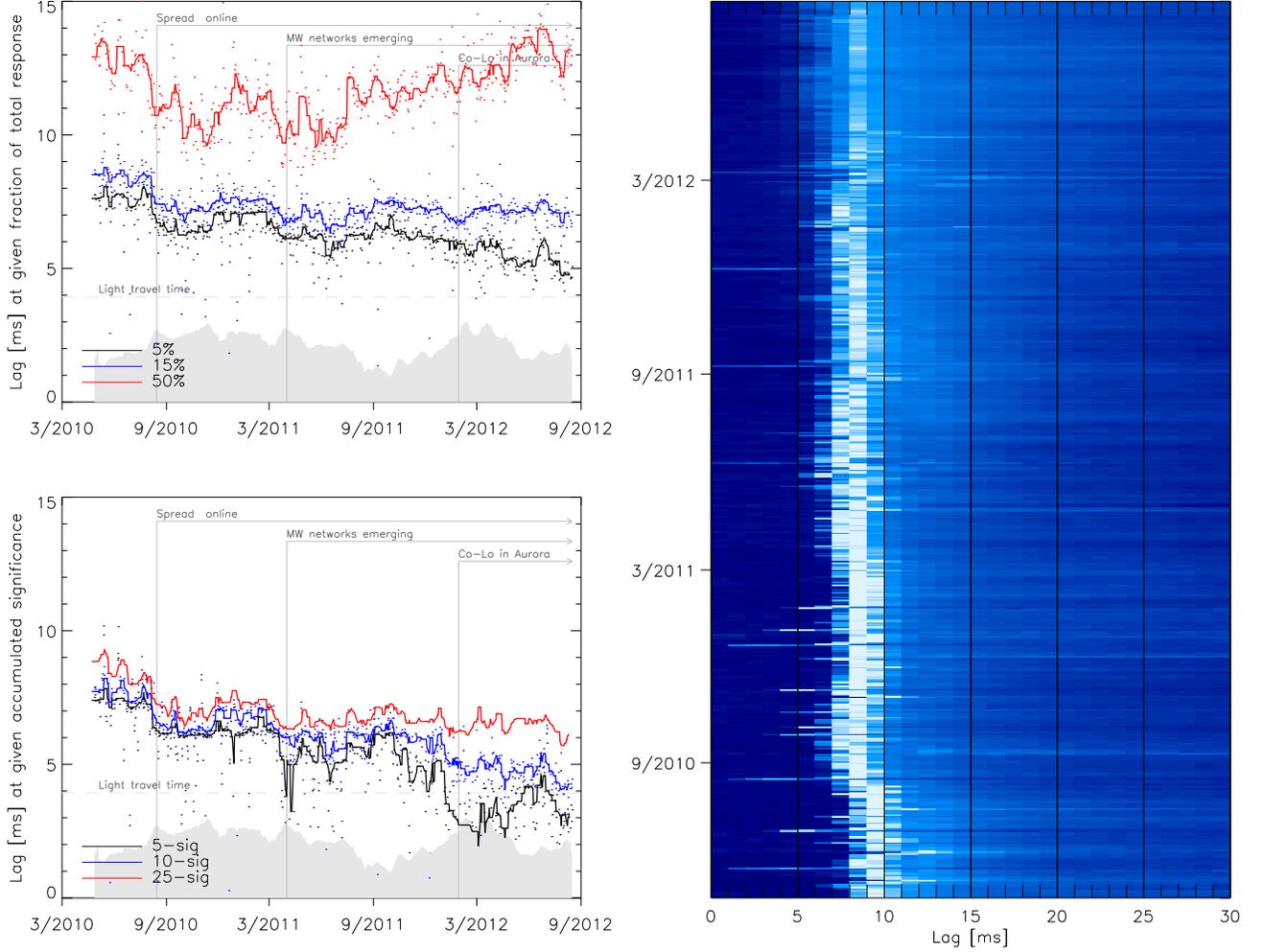}
\caption{ \footnotesize{{\it Upper left}: Evolution of the $X=5$\%, $X=15$\%, and $X=50$\% liquidity response estimates -- see text in Sec.~\ref{sec-measure} for definitions. Curved are obtained via 10-day median filtering of individual days (points). {\it Bottom left}: Evolution of the accumulated significance measures, $t(5\sigma)$, $t(10\sigma)$, and $t(25\sigma)$ for the liquidity response. {\it Right}: All of the volume-normalized liquidity response curves from the $N=600$ analyzed days in the form of a heat map. The intensity scale runs from dark blue (zero response) to white (maximum response). The shaded region in the left panel shows the relative total integrated response $T$ for each day, by which the daily response is normalized in the right panel.}}
\label{fig-trade-liq-ev}
\end{figure*}

\begin{figure*}[ht]
 \centering
 \includegraphics[width=18cm,angle=0]{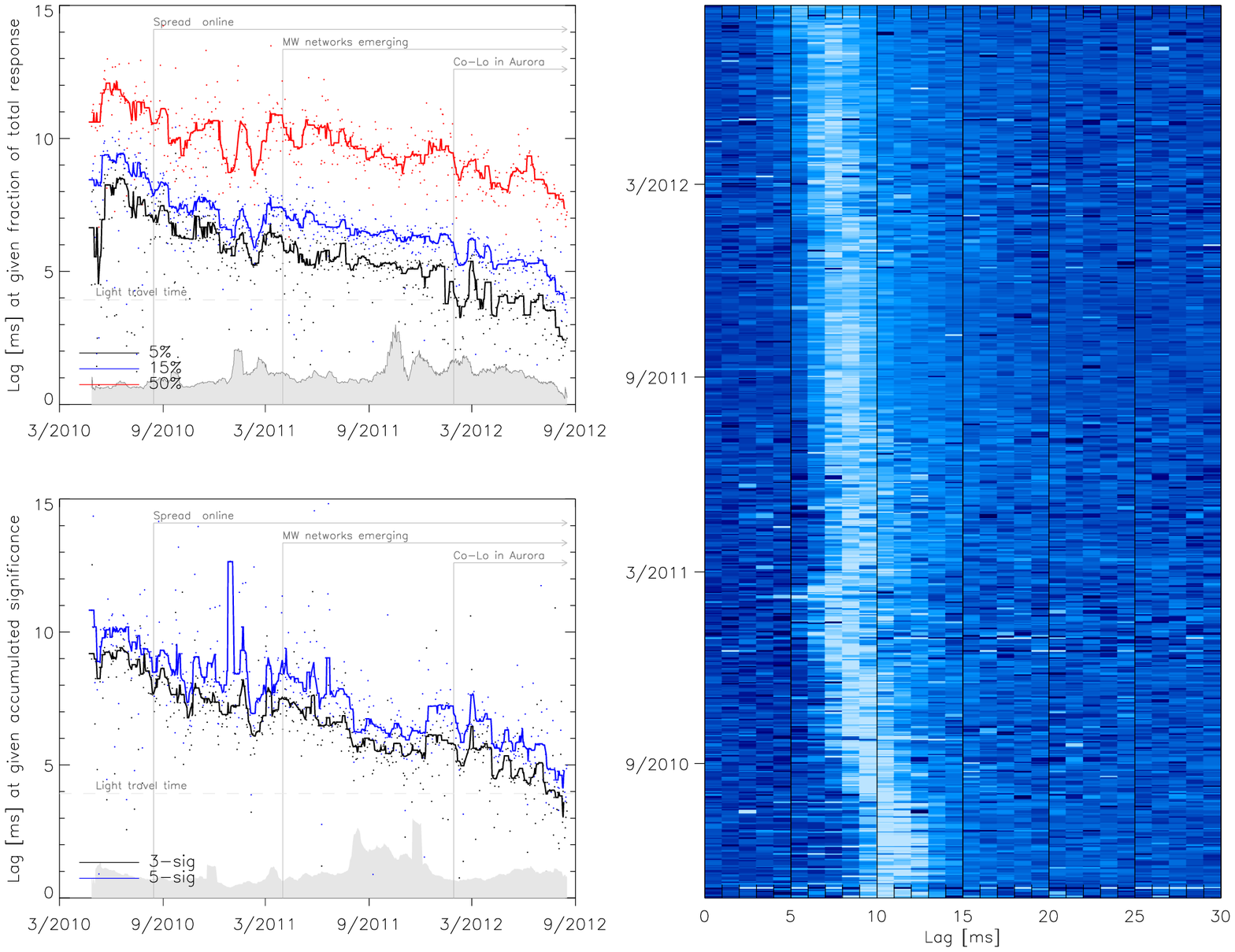}
\caption{ \footnotesize{{\it Upper left}: Evolution of the $X=5$\%, $X=15$\%, and $X=50$\% trade response estimates -- see text in Sec.~\ref{sec-measure} for definitions. {\it Bottom left}: Evolution of the accumulated significance measures, $t(3\sigma)$, and $t(5\sigma)$ for the trade response. {\it Right}: All of the volume-normalized liquidity response curves from the $N=600$ analyzed days in the form of a heat map. The intensity scale runs from dark blue (zero response) to white (maximum response). The shaded region in the top left panel shows the relative total total trade response {\em in dollars} for the period following the CME trades; that in the bottom left shows the total integrated trade response $T$ for each day, by which the daily response is normalized in the right panel.}}
\label{fig-trade-trade-ev}
\end{figure*}

To some degree, some of these effects could in principle be partially disentangled using additional levels of analysis, and/or using data that is not available to us (such as the identity of firms originating each order, or a data feed that is independently time stamped at the exchange gateway.) For purposes of this paper we have settled on two sets of measures. The time evolution of the statistics $t_{l\,X}$ and $t_{t\,X}$ as defined in Sec.~\ref{sec-measure} can be used to sample the market evolution. The upper-left panels of figures \ref{fig-trade-liq-ev} and \ref{fig-trade-trade-ev} show these quantities for $X=5$\%, $X=15$\%, and $X=50$\% for the trade-liquidity response and the trade-trade response respectively.  We compute a second latency measure by computing, for time $t > t_0$, the accumulated significance of a detectable response: we compute both the integrated response $T(t)$ from $t_0$ to $t$, and the uncertainty $\sigma_T$ in this quantity, computed by summing the $1\sigma$ uncertainties in quadrature for the bins at earlier times.  We then take $T/\sigma_T$ as a measure of response significance; and (via linear interpolation) determine the time $t(\sigma)$ at which this significance reaches a number $\sigma$ of standard deviations.  These times are plotted in the lower-left panels of figures \ref{fig-trade-liq-ev} and \ref{fig-trade-trade-ev}, with $t(3\sigma)$ and $t(5\sigma)$  plotted for the trade response, and 5, 10, and $25\sigma$ responses plotted for the liquidity data.  This measure gives both insight into the latency evolution, and also an indication of the per-day statistical significance of detecting a signal at a given latency.
 
Examining these quantities reveals a substantial day-to-day variation in all of them; this variation greatly exceeds the statistical error on the data points (particularly in the liquidity response, which has small uncertainties), and likely points to shifting players, strategies, matching engine load at the CME, and other factors.  Nonetheless, clear trends in the response can be discerned, and these can be compared with known external events affecting market response times. First, there is an overall downward trend in all ``leading edge" indicators, namely $t_{l\,5}$, $t_{t\,5}$, $t(3\sigma)$ for trades, and $t(5\sigma)$ for liquidity.  In agreement with the results in Fig.~\ref{fig-trade_trade_resp_curves}, these trends hold also for $t_{t\,15}$ and $t_{t\,50}$, indicating that the overall shape of the trade response is relatively constant, while its response time has steadily shifted to lower latency.  In contrast, while the leading edge of the liquidity signal has shifted to lower response time, the $t_{l\,50}$ has {\em increased}, indicating that the response has significantly broadened even while the leading edge has shifted to lower latency.

During the time of our data sample, a number of known external events potentially affecting latency occurred; some of these are indicated in Figures~\ref{fig-trade-liq-ev} and~\ref{fig-trade-trade-ev}. Prior to summer/fall 2010, the fastest communication route connecting the Chicago futures markets (with point of presence at Equinix's 350 E. Cermak data center) to the New Jersey equity markets allowed a response 7.25-7.95 ms after a price change in Chicago~\cite{Adler}. This is in accord with Figures~\ref{fig-trade-liq-ev} and~\ref{fig-trade-trade-ev}, which show first responders in the 7 and 8 ms bin.

In August of 2010, Spread Networks introduced a new custom fiber optic line that was shorter than the pre-existing route and used lower latency equipment. This technology reduced latency between 350 Cermak and  Verizon's Carteret Data Center that houses the NASDAQ matching engine, with the lowest-latency service running at a quoted 6.65\,ms (recently reduced to 6.55\,ms.)  In Figures~\ref{fig-trade-liq-ev} and~\ref{fig-trade-trade-ev}, one can discern a latency drop of $\approx 1-1.5\,$ms near this time.  Note that around this time, the CME group moved its matching engine from 350\,Cermak to its new Aurora data center, presumably adding a sub-ms latency between this matching engine and the NJ data centers (in principle, the fiber communication time between these points can be $< 250\, \mu s$, but we have no data as to the actual latency introduced.)

According to FCC records, in Fall 2010 the first clandestine MW route became fully licensed.  A MW route with expedited construction can probably be assembled in 6-12 months after the licensing process begins.  Under this running (and very approximate) assumption, one MW route may have been operating in early 2011, followed by several more in mid/late 2011, and up to a half-dozen as of spring/summer 2012.  Each of these routes may serve one or more trading firms.  In Figures~\ref{fig-trade-liq-ev} and~\ref{fig-trade-trade-ev} there is a gradual decrease in latency, with the first response dropping from 6-7\,ms in Fall 2010 to 3-4\,ms as of August 2012.  We attribute this steady latency decrease to the construction, commissioning, and adoption of microwave data transmission by numerous firms during this period. 

Also of note is a possibly significant downward latency drop in late January 2012.  On January 29, 2012, CME-provided co-location moved from 350 Cermak to the Aurora data center.  For MW paths terminating in Aurora, this potentially cut out a round-trip $\lesssim 0.5\,{\rm ms}$ path from Aurora (where the matching was occurring) to 350 Cermak (where CME allowed colocation) back to Aurora (where the MW network began).   
Another significant decrease occurred in early July 2012.  It is possible that this is connected with the launch of a particular network used by multiple players that was able to significantly shift the response time.

\section{Estimated Costs and Relation to US Telecommunications Infrastructure}

\begin{figure*}[ht]
 \centering
 \includegraphics[height=4.5cm,angle=0]{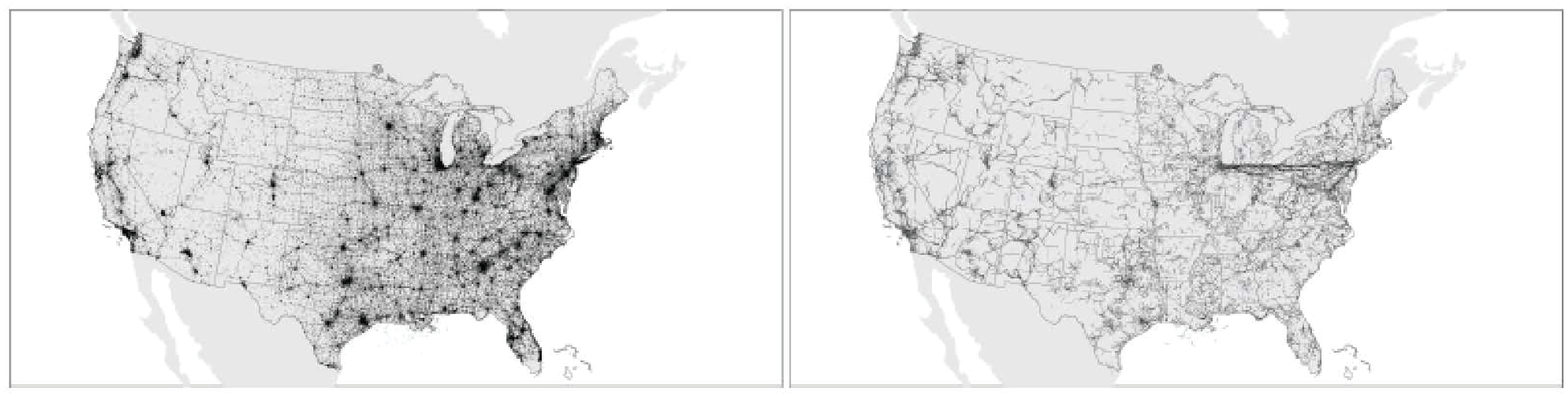}

\caption{A map of 93,600 paths present in 187,338 microwave site-based license applications for the two years prior to September 1, 2012 (left) and of 191,290 towers culled from the FCC database (right). Colors indicate frequency bands: 6 GHz (black), 11 GHz (charcoal), 18 GHz (dark gray), and other (slate gray).
\label{fig-maps}}
\end{figure*}

The development of Spread Networks' fiber link is publicly estimated to have cost approximately \$300 million~\cite{Steiner}, exclusive of substantial operating costs, both for Spread Networks and for the users; for example, traders who lease dark fiber pairs incur an additional cost of lighting their link so that it can convey information between the markets.

We can roughly estimate the capital and recurring expense of the ongoing MW network effort as follows.  Turnkey installation of a commercial MW hop, including equipment, path engineering, site and path surveys, frequency coordination and licensing, tower work, installation, and commissioning, total of order \$100,000-\$250,000. Because this market segment can involve customized equipment and more proprietary capabilities than for more general-purpose MW builds, costs are likely to be closer to the high end of this estimate.  Given a total of 554 links in Table~\ref{tab-routes}, this suggests a CapEx expense significantly exceeding $\sim \$139$ million; this does not include a number of networks in less-advanced stages of their build, networks that may not show up in the FCC database, or other costs.

Operating costs for a MW network can be estimated starting with the generally dominant cost of tower lease costs. A ``rule of thumb" for such costs in this region is $\sim $\$100/vertical antenna foot/month. (Mike Hunter, RCC Consultants, Private Communication).  Assuming two 8-foot antennas per link, the 554 links in Table~\ref{tab-routes} represent $\sim\,$\$10$\,{\rm m\, yr^{-1}}$.  Other operating costs such as power, maintenance, network operating centers, etc., are likely to roughly double this cost.  Thus over five years OpEx adds approx. \$100 million, for a total of approx. \$250 million over five years for the MW effort even if no additional networks are deployed and networks outside of the NY-Chicago corridor are neglected.

An interesting implication of these estimates as combined with the findings of Sec.~\ref{sec-mwnetworks} is that lower-latency MW networks are generally {\em less} expensive than slower networks, because the prime factor in both latency and cost for current networks is the number of hops.  For example, at \$250,000/hop, there is a range from \$4.5m-\$20m in the approximate cost of currently-licensed networks, but the fastest networks are at the low end of the spectrum, and the slowest at the high end; similar considerations apply for operating expenses as well.  Thus as well as the latency, the cost of networks has been largely determined by the skill in their planning.  It also implies that future networks competing or surpassing the current best could cost relatively little in capital expense, most likely $\lesssim\, $\$5-8\,m.

The MW networks being deployed for financial data transmission have by some measures become a non-negligible component of the U.S. wireless network industry. Figure~\ref{fig-maps} shows a map of the two years prior to September 1, 2012, in microwave site-based FCC applications.  The Chicago-New York-Washington triangle is clearly visible.  During this period, approximately 94,000 paths totaling 1.3 million km were applied for;\footnote{Note that these figures count each direction and each frequency of a MW link separately; the number of applications can exceed the number of paths for various reasons including revisions, updates, etc.} of this, at least 1422 paths totaling $\sim$ 51,606 km appear related to financial data networks on the Chicago-NJ route.  The Chicago-NJ networks utilize approx. 490 unique towers, from of order 190,000 that appear in the FCC database, 19,000 of which lie within the 2-degree latitude swathe running between Chicago and NJ.   The effort has also spurred significant technological investment.  On the basis of the FCC and other data we estimate that in the space of two years, at least thirteen MW radio manufacturers have developed customized equipment aimed at the low-latency segment.

This infrastructure has presumably been motivated by greater profits accessible to traders with access to lower latency links. We can use our data sets to make an example point estimate of the relationship between returns and latency by examining the record of SPY trading during the April 27th, 2010 -- August 17th, 2012 time frame. Changes in the traded price of the near-month E-mini contract at the CME generate substantial (and accurately measurable) increases in trading volume for the SPY ETF. On average, during the full period spanned by our analysis, 15\% of all SPY trades (amounting to $V_{\rm CME \rightarrow SPY}=20\,$ billion shares) can be specifically identified as excess volume associated with the HFT response to E-mini price moves. The trading rate for SPY increases by a factor of $\sim30\times$ when the E-mini ticks up or down. Sharp increases in SPY trading volume begin when the news of an E-mini price changes reaches the New Jersey exchanges, and the rate of excess trading displays a time-dependent behavior that closely tracks the trade-trade response functions discussed in \S 2. On average, during the period of our analysis, the E-mini was priced at \$65,000, with a \$12.5 bid-offer spread, whereas shares of SPY were priced at \$130, with a \$0.01 minimum spread. Given that the two instruments are fungible into each other, a one-tick move for the E-mini corresponds to a $(1.25\times10^{1}/6.5\times10^{4})\times \$130 = \$0.025$ movement in SPY. Therefore, the total PnL, ${\cal P}_{\rm SPY}$, {\it specifically attributable} to latency improvement for the $V_{\rm CME \rightarrow SPY}$ traded shares is of order ${\cal P}_{\rm SPY}\sim (2.0\times10^{10}\,{\rm sh.})\times\$0.025\,{\rm sh.}^{-1}=\$0.5\,$ billion, a figure that, interestingly, is of the same order as the investment in low latency infrastructure between Chicago and New Jersey. (These estimates can be extended to other equities with identical methodology, but this data is not reported here.) Using our data, we can calculate, for a given trading period, the one-way latency, $L_{f}$ required to capture a given fraction, $f$, of ${\cal P}_{\rm SPY}$. We find, for example, that during the period covered by our analysis, $L_{0.95}$ has dropped from $L_{0.95}=8\,{\rm ms}$ to $L_{0.95}=5\,{\rm ms}$.  

Positional competition at the sub-millisecond level is, however, more complex than this simple example suggests. Although algorithms will, in general, yield higher returns at lower latency, the tradeoff between return and latency is likely a complex function of the algorithm, competing algorithms, and the spread of latencies among competitors.
 
\section{Discussion}

The evolution of the inter-market latency between the Chicago Mercantile Exchange and the New Jersey-based equity exchanges serves as an interesting example of the development of ever-faster infrastructure for securities trading. The speed of information transport between Chicago and New York has decreased significantly during the past three years, which we attribute to the launch of an optimized fiber optic link, the emergence of line-of-sight microwave networks connecting the financial centers, and the increased presence of trading algorithms in the equity markets that are able to either pre-coordinate orders across exchanges or to anticipate price movements in the futures markets. In the latter case, the coordination of trades across different exchanges is already a technological challenge; the incredibly low latencies that may evolve over coming years may also quite soon start to involve higher-order relativistic effects, and challenge even the GPS system's ability to define and maintain a sufficiently accurate globally synchronized time frame required for such coordinated orders to be effective.  

We have found that the latency measurements derived from our analysis of exchange-provided financial tick data are broadly consistent with the inferred properties of the licensed microwave networks that appear in publicly available FCC records. Our analysis, furthermore, provides strong support for the hypothesis that price formation for the United States equity market occurs in the near-month E-Mini S\&P 500 Futures contract. As a consequence, changes in the traded price of the E-Mini contract provide a continuous sequence of small exogenous shocks, to which the US equity markets respond in a time-averaged manner that is both predictable over a timescale of order $\tau_{\rm lat}=10\,{\rm ms}$ and is highly statistically significant.  Our identification of this predictability, along with the demonstration that the CME acts as a drive system with the equity exchanges acting as response systems, raises the intriguing possibility that ideas related to the synchronization of chaos and Lyapunov characteristic exponents (see, e.g.~\cite{Bezruchko}) might have applicability to the characterization of market behavior. In this regard, we speculate that ``dissipation'' in a market viewed as a dynamical system can be identified with the noise trading provided by retail orders, and that the Lyapunov time,  $\tau_{L}$, is of order $\tau_{L} \sim \tau_{\rm lat} \sim10\,{\rm ms}$, as measured by the exponential decay observed in our response curves over this time scale. If this view has merit, then the equity exchanges display a dramatic contrast in their Lyapunov time scale to that shown by Earth's weather, which exhibits predictability over periods measured in days \cite{Lorenz}, and to Earth's orbital motion, which can be predicted accurately for millions of years, but which becomes completely unknowable on time scales exceeding $\tau\sim 100\,{\rm Myr}$~\cite{Laskar}. 

It is remarkable that -- as demonstrated in this paper -- an appreciable fraction of the entire U.S. equities market responds, about once per second, to just a few bits of information emanating from suburban Chicago and traveling via various channels and between 4-10 milliseconds, to suburban New Jersey. This fact makes the ability to access those bits as fast as possible worth ten or hundreds of millions of dollars per year.  This is perhaps an unusual case, resulting from the geographic separation of the dominant U.S. futures and equities markets.  Nonetheless the expenditures being made to reduce latency over other international financial centers. For example, the 7,800 km Asia Submarine Cable Express, linking the Singapore and Tokyo markets, recently opened. It reportedly cost  approximately \$430 million and reduces latency over that route by 3 ms~\cite{Grant}. A similar $\sim$\$300 million, 6021\,km cable project underway by Hibernia will reduce the New York-London latency by $\sim 6\,$ms \cite{TelegraphArticle},
and a trans-polar fiber route, with reported estimated cost of \$1.2\,billion has been launched that will reduce the Asia-Europe communication time by tens of milliseconds\cite{Guardian}
indicate that similar financial incentives exist on various routes worldwide.  

This has led to an interesting physics and technology problem, of how a given bit rate of information can be communicated as quickly as possible between widely separated points.  The current answer appears to be via microwave networks overland, and via fiber optic cables undersea.  How this will evolve in the future depends on a number of factors, including those in the regulatory space not addressed here.  In the short term, it is likely that MW and fiber networks will continue to be further latency-optimized (perhaps, for example, including hollow-core fibers in which information travels at the vacuum speed of light.)  Other, physically well-understood but less robust technologies such as floating microwave stations or drones may come into play.  In the distant future, we can speculate that exotic technologies such as neutrino or even WIMP, axion or gravity wave communications could be employed to communicate financial tick data directly through the Earth.

\section*{\small{Acknowledgments}}
\small{We are grateful to Darrell Duffie, Graham Giller, Neil Miller, Philip Nutzman, Matt Simsic, Max Tegmark, and Jaipal Tuttle for useful discussions and constructive comments on the manuscript. We thank David Levine for researching microwave tower availability rates. We thank Zachary Maggin of NASDAQ OMX for assistance with obtaining the historical ITCH4.1 tick data described herein on an academically fee-waived basis. We thank Matt Frego of the CME for processing our Data Mine Academic order for historical data, as well as for our ongoing standard Data Mine orders. We thank the Rock Center for Corporate Governance at Stanford University for funding and research support.}

\end{document}